\documentstyle[aps,prl,epsf,multicol]{revtex}

\begin{document}
\title{On the distribution of barriers in the spin glasses}
\author{L. B. Ioffe}
\address{Physics Department, Rutgers University, Piscataway, NJ 08855 \\
and Landau Institute for Theoretical Physics, Moscow}
\author{D. Sherrington}
\address{Theoretical Physics, 1 Keble Rd, Oxford OX1 3NP, UK}
\maketitle

\begin{abstract}
We discuss a general formalism that allows study of transitions over
barriers in spin glasses with long-range interactions that contain large but
finite number, $N$, of spins. We apply this formalism to the
Sherrington-Kirkpatrick model with finite $N$ and derive equations for the
dynamical order parameters which allow ''instanton'' solutions describing
transitions over the barriers separating metastable states. Specifically, we
study these equations for a glass state that was obtained in a slow cooling
process ending a little below $T_{c}$ and show that these equations allow
''instanton'' solutions which erase the response of the glass to the
perturbations applied during the slow cooling process. The corresponding
action of these solutions gives the energy of the barriers, we find that it
scales as $\tau ^{6}$ where $\tau $ is the reduced temperature.
\end{abstract}

%\newpage

\begin{multicols}{2} 

The most prominent feature of a glass state of matter is the existence of an
extensive number of metastable states separated by large energy barriers.
This feature is reproduced in many infinite range models which allow a mean
field treatment; in the framework of this mean field approach the properties
of these states have been studied extensively over the last 20 years and a
very detailed picture has emerged \cite{SGbooks}. Originally, the models
studied were thought to describe only systems with frozen disorder, like
spin glasses, but more recently it was realized that similar methods can be
applied to frustrated systems without quenched disorder that may be viewed
as analogues of ordinary glasses \cite{Feigelman97,CuKu97}.

Empirically, the existence of the extensive number of states is revealed in
a very slow dynamics of the glass phase that is referred to as creep or
ageing; this slow dynamics is usually attributed to transitions between
different metastable states. Although, as a result of this extensive
theoretical work mentioned above, the properties of the individual states
are well understood, transitions between these states have been deprived of
due attention. In this Letter we propose an analytical approach to address
this problem; we apply it to the Sherrington-Kirkpatrick model \cite
{Sherrington} and derive integral equations which describe the transitions
between metastable states. We were not able to solve these equations
analytically but we can show numerically that they admit a solution with
expected qualitative properties. Our results have two applications: firstly,
we hope that most qualitative features of the barrier distribution found in
the infinite range models will hold in the finite range physical glasses,
and secondly, it seems possible to test the predictions of the infinite
range models on purpose built Josephson arrays \cite{Gersheson96}; in the
latter systems the number of effective ``spins'', $N$, is large ($N\sim
100-300$) but not infinite and our results should be directly applicable.

Specifically, we shall consider a dynamical version of the
Sherrington-Kirkpatrick model with $N$ soft spins that is characterized by
the equations of motion 
\begin{eqnarray}
\Gamma _{0}^{-1}\partial _{t}S_{i}(t) &=&\frac{\delta (\beta H)}{\delta
S_{i}(t)}+\zeta _{i}(t)  \label{dS/dt} \\
H &=&-\sum_{i,j}J_{ij}S_{i}S_{j}-\sum_{i}S_{i}h_{i}+\sum_{i}V(S_{i})
\label{H}
\end{eqnarray}
Here $J_{ij}$ is a matrix of quenched random Gaussian couplings normalized
by $\langle J_{ij}^{2}\rangle =1/N$, $h_{i}$ is an external magnetic field
acting on spin $i$, $\zeta (t)$ is random thermal noise with correlator $%
\langle \zeta _{i}(t)\zeta _{j}(t^{\prime })\rangle =2\Gamma _{0}^{-1}\delta
_{ij}\delta (t-t^{\prime })$ and $V(S)$ is a one spin potential that keeps $%
\langle S_{i}^{2}\rangle =1$; its exact form is irrelevant, but for example
one can take $V(S)=r_{0}(S^{2}-1)^{2}$, with $r_{0}\gg T$.

It is well established that the low temperature state of this model is
completely characterized by two order parameters: the correlation function, $%
D(t_{1},t_{2})=\langle S_{i}(t_{1})S_{i}(t_{2})\rangle $, and the response
function, $G(t_{1},t_{2})=\langle \frac{\delta S_{i}(t_{1})}{\delta
h_{i}(t_{2})}\rangle $. In the low temperature state both these functions
acquire a part which is non-decaying in time \cite{Sompolinsky82}; a
particular feature of the glass phase is the appearance of a non-decaying
contribution to the response function $G(t_{1},t_{2}),$ showing that a
perturbation applied in the distant past (at time $t_{2}$) continues to
affect the present (at time $t_{1}$). Furthermore, different sample
histories leading to the same final temperature and magnetic field would
produce different order parameters, $D(t_{1},t_{2})$ and $G(t_{1},t_{2})$ 
\cite{Vinokur87,Ioffe88}; this is what one would expect in a system that can
be trapped in any of many metastable states and where the state where it is
trapped is determined by the thermal history. So, in order to specify the
glass state one needs to specify the history of the sample that led to this
state. In the following we shall consider the systems prepared by the slow
cooling process in which the temperature is varied slowly compared to the
spin flip rate, $\Gamma _{0}$. In this regime it is convenient to separate
the correlation and response functions into ``fast'' and ``slow'' parts: $%
D(t_{1},t_{2})=D_{f}(t_{1},t_{2})+q(t_{1},t_{2})$ and $%
G(t_{1},t_{2})=G_{f}(t_{1},t_{2})+\Delta (t_{1},t_{2})$. The order
parameters characterizing the state of the model at the end of a monotonic
slow cooling process ending at $T(t_{1})=T_{1}$ are well known \cite
{Vinokur87,Ioffe88} if the reduced temperature is small, $%
(T_{c}-T)/T_{c}=\tau \ll 1$: 
\begin{equation}
\begin{array}{ll}
q_{t_{1},t_{2}}=\min (\tau _{1},\tau _{2})+O(\tau ^{2}) &  \\ 
\Delta (t_{1},t_{2})=\theta (t_{1}-t_{2})2\tau _{2}\left( \frac{d\tau }{dt}%
\right) _{t_{2}} &  \\ 
\delta q_{t}\equiv \tau _{t}-q_{t,t}=-\tau _{t}^{2} & 
\end{array}
\label{q_ad}
\end{equation}
Here and in the following we denote $\tau _{1}\equiv \tau (t_{1})$. This
form of $\Delta (t_{1},t_{2})$ shows that any perturbation applied during
the cooling process has everlasting effects; this is natural since
qualitatively each state continuously subdivides during the cooling process
and no transitions between these states are possible within the framework of
this approach which becomes exact at $N\rightarrow \infty $ and $\frac{dT}{dt%
}\rightarrow 0$. Empirically, we expect that at finite $N$ these transitions
become allowed the response function $\Delta (t_{1},t_{2})$ decays and
eventually the memory of a perturbation applied at a past time is lost.
Furthermore, we expect that barriers formed between states with larger
overlap are generally smaller than the barriers between states with small
overlap and that transitions between the former are therefore more likely to
occur. States with large overlaps are formed at later stages of a cooling
process as a subdivision of the ancestor state, so we expect that the lowest
barriers would correspond to the transitions which destroy only the memory
of a recent past.

In fact, as we shall show below, a theory in terms of only $q$ and $\Delta $
is insufficient to describe transitions over the barriers and we need to
introduce one more order function. We explain the general formalism using
the simple example of a particle in a potential $V(x)=vx^{2}(1-x)$ which is
in the metastable state around $x=0$ at low temperatures $T\ll v$ but might
escape to $x=\infty $ with probability $\exp (-V_{max}/T)$. All correlation
functions in this warm-up problem can be obtained from the dynamic equations
of motion 
\begin{equation}
\frac{dx}{dt}=-\frac{dV}{dx}+\zeta (t),\;\langle \zeta (t)\zeta (t^{\prime
})\rangle =2T\delta (t-t^{\prime })  \label{dx/dt}
\end{equation}
or, equivalently, from the generating functional 
\begin{eqnarray}
Z \!&=&\!\int {\cal D}x{\cal D}\hat{x}\exp (-{\cal A}\{x,\hat{x}\}+{\cal A}%
_{h}\{x,\hat{x}\}){\cal J}\{x\}  \label{Z} \\
{\cal A}\{x,\widehat{x}\} &=&\int \left[ i\hat{x}(\frac{dx}{dt}+\frac{dV}{dx}%
)+T\hat{x}^{2}\right] dt  \label{Ax}
\end{eqnarray}
where ${\cal J}\{x\}=\det \left( \frac{d}{dt}+\frac{d^{2}(\beta V)}{dx^{2}}%
\right) $ is a Jacobean whose effects one can ignore to get the results with
only an exponential accuracy and ${\cal A}_{h}\{x,\hat{x}\}$ are source
terms, e.g. in order to get correlators of $x$ one can use ${\cal A}_{h}\{x,%
\hat{x}\}=hx$, then $\left\langle x^{k}\right\rangle =\frac{d^{k}Z}{dh^{k}}$%
. At small temperatures the functional integral in (\ref{Z}) is dominated by
the saddle point solutions. Varying the action ${\cal A}\{x\}$ with respect
to $x$ and $\hat{x}$ we get 
\begin{equation}
\begin{array}{ll}
\frac{d\hat{x}}{dt}-\frac{d^{2}V}{dx^{2}}\hat{x}=0 &  \\ 
\frac{dx}{dt}+\frac{dV}{dx}=2iT\hat{x} & 
\end{array}
\label{x_eq}
\end{equation}
These equations admit two dramatically different solutions. In the first
solution $\hat{x}=0$, $\frac{dx}{dt}=-\frac{dV}{dx}$ corresponding to a
particle that slides down the potential hill practically unaffected by the
thermal noise; this solution has zero action. The second solution is $iT\hat{%
x}=\frac{dx}{dt}$, $\frac{dx}{dt}=\frac{dV}{dx}$, corresponding to the
particle going up the hill instead of down; it can be obtained from the
first solution by inverting the sign of the time so that the response of
this particle to an external field is anti-causal. Within the saddle point
approximation the path of the particle escaping the metastable state at $x=0$
consists of the latter solution for $0<x<x_{max}$ and the former for $%
x>x_{max}$ (here $x_{max}$ is the point where potential has a maximum); the
action associated with this path is $A=\beta \delta V$, so it has
probability $exp(-\beta \delta V),$ reproducing, as it should, the Boltzmann
formula. Note that the general statement that the correlator $\langle \hat{x}%
\hat{x}\rangle =0$ does not prevent one from having this ``instanton''
solution with a finite conjugate field. Note also that once the solutions
with this field non-zero are allowed, the action corresponding to these
solutions is no longer zero but this does not contradict the general
statement that $Z\equiv 1$ in the absence of source terms in (\ref{Z})
because Jacobean contribution cancels the contribution of these solutions in
the absence of source terms. As a more detailed study \cite{Feigelman} shows
in the presence of the source terms Jacobean contribution does not affect
the leading exponential factor and that therefore the simple analysis (like
the one above) that ignores source terms and the effects of Jacobean gives
correct result for correlators. In what follows we shall assume that this
conclusion is also true for the glass problem and ignore the effects of
Jacobean and of the source terms.

A similar generating functional procedure with extremal analysis is
conveniently employed to discuss the spin glass models of Eqs. (\ref{dS/dt})
and (\ref{H}). By analogy with above simple example we expect that saddle
point solutions corresponding to transitions over the barriers exist also in
the glass model if the fields conjugate to the variables $D(t_{1},t_{2})$
and $G(t_{1},t_{2})$ are allowed to acquire non-zero values. So, we have to
repeat the derivation of the equations for the order parameters without
making the usual assumption that the variables conjugated to $D(t_{1},t_{2})$
and $G(t_{1},t_{2})$ are zero. The first steps of the derivation are not
changed; these steps are explained well in the literature \cite
{Sompolinsky82,Vinokur87,Ioffe88,Kurchan93}, so we only recall them here.
The dynamics (\ref{dS/dt}) is reproduced by the generating functional $%
Z=\int {\cal D}S{\cal D}\hat{S}{\cal D}\bar{\chi}{\cal D}\chi \exp (-{\cal A}%
_{B}\{S,\hat{S}\}-{\cal A}_{F}\{\chi ,\bar{\chi}\})$ where $S$, $\widehat{S}$
and ${\cal A}_{B}\{S,\hat{S}\}$ are analogous of $x$, $\widehat{x}$ and $%
{\cal A}_{{}}\{x,\widehat{x}\}$ and additional integrals over fermionic
fields $\chi ,\bar{\chi}$ reproduce the Jacobean. To simplify the notations
it is very convenient to introduce, following Kurchan \cite{Kurchan93},
supercoordinates $\theta $ and $\bar{\theta}$ that allow one to write the
action in terms of the superfield $\phi (\theta ,\bar{\theta})\equiv
S+\theta \bar{\chi}+\bar{\theta}\chi +\bar{\theta}\theta \hat{S}$. In these
compact notations the full action ${\cal A}\{\phi \}$ can be represented as
a sum of a single spin part, ${\cal A}_{0}$, and interacting part, ${\cal A}%
_{int}=i\frac{1}{2}\sum_{i,j}J_{ij}\phi _{i}\phi _{j}$, the latter being the
only term containing the couplings $J_{ij}$. This action is linear in the
coupling $J_{ij}$, allowing one to average the generating functional over
the Gaussian distribution of $J_{ij}$. This results in a four `spin'
interaction which one can decouple by a symmetric superfield $Q_{{\cal T}%
_{1},{\cal T}_{2}}$ (here and below we use notation ${\cal T}$ to denote the
set of variables $\{t,\theta ,\bar{\theta}\}$. As can be shown \cite
{Feigelman} in the illustrative example discussed in the last paragraph, we
assume that the fermionic contribution can be ignored in a study to leading
(exponential) order, so we retain only Bose components of the order
parameter field $Q_{{\cal T}_{1},{\cal T}_{2}}$ : 
\begin{equation}
\begin{array}{ll}
Q_{{\cal T}_{1},{\cal T}_{2}}^{(B)}=D_{t_{1},t_{2}}+G_{t_{1},t_{2}}\bar{%
\theta _{2}}\theta _{2}+\tilde{G}_{t_{1},t_{2}}\bar{\theta _{1}}\theta _{1}+
&  \\ 
\hat{D}_{t_{1},t_{2}}\bar{\theta _{1}}\theta _{1}\bar{\theta _{2}}\theta _{2}
& 
\end{array}
\label{Q}
\end{equation}
The symmetry $Q_{{\cal T}_{1},{\cal T}_{2}}=Q_{{\cal T}_{2},{\cal T}_{1}}$
implies that functions $D_{t_{1},t_{2}}$ and $\hat{D}_{t_{1},t_{2}}$ are
symmetric and that $\tilde{G}_{t_{1},t_{2}}=G_{t_{2},t_{1}}$, so $\tilde{G}%
_{t_{2},t_{1}}$ is a backward response. The next technical step is to
integrate out the spin variables which are now decoupled and get the
effective action of the field $Q({\cal T}_{1},{\cal T}_{2})$: 
\begin{eqnarray}
Z &=&\int {\cal D}Q\exp (-N{\cal A}\{Q\})  \label{Z_Q} \\
{\cal A} &=&\int \frac{1}{4}\beta _{t_{1}}Q_{{\cal T}_{1},{\cal T}%
_{2}}^{2}\beta _{t_{2}}d{\cal T}_{1}d{\cal T}_{2}+{\cal A}_{S}\{Q\}
\label{A}
\end{eqnarray}
where 
\begin{eqnarray}
{\cal A}_{S}\{Q\} &=&-\ln \left[ \int {\cal D}\phi \exp (-{\cal A}_{0}\{\phi
\}+\right. \\
&&\left. \frac{1}{2}\int \beta _{t_{1}}Q_{{\cal T}_{1},{\cal T}_{2}}\beta
_{t_{2}}\phi ({\cal T}_{1})\phi ({\cal T}_{2})d{\cal T}_{1}d{\cal T}%
_{2})\right]  \label{A_S}
\end{eqnarray}
is the effective action of $Q$ generated by decoupled spin degrees of
freedom, $\phi $. The large factor $N$ in front of the action in Eq. (\ref
{Z_Q}) allows one to look only for the saddle point solutions which are
obtained by varying the action with respect to $Q$ or, more explicitly, with
respect to its components $\hat{D}$, $D$ and $G$. The solution with zero
action corresponds to the usual choice $\hat{D}_{t_{1},t_{2}}=0$, $%
G_{t_{2},t_{1}}=0$ at $t_{1}>t_{2}$; but neither of these relations
necessarily holds for the saddle point solution with non-zero action that we
are looking for. As we shall show explicitly below, if these conditions are
imposed the usual solutions for the response functions of the glassy state
are recovered. We shall of course need an explicit form of ${\cal A}%
_{S}\{Q\} $; in the general case this action is very complicated, so to
simplify we shall consider only the vicinity of the transition temperature ($%
\tau \ll 1$) and keep only the terms which affect the long time dynamics at
such temperatures. Furthermore, we shall keep only the leading terms in the
expansion in the conjugate order parameters $\hat{D}_{t_{1},t_{2}}$ and $%
G_{t_{1}<t_{2}}$ because, as we shall show below, transitions over small
barriers (which, of course, dominate) involve only small values of these
quantities.

Below we shall use a diagram expansion to derive the full saddle point
equations following from the action (\ref{A}). However, before we embark on
it, it is instructive to do a preliminary analytic calculation assuming that
one can keep only quadratic terms in the action ${\cal A}_{0}\{\phi \}$. The
results of this calculation can not be used below $T_{c}$, because there
non-linear terms are essential but this derivation is much simpler and gives
correctly all terms of the full equations except one thereby providing a
useful reference point. In this approximation all integrals over $\phi $ in (%
\ref{A_S}) are Gaussian and can be performed explicitly, so ${\cal A_{S}}$
becomes: 
\begin{equation}
{\cal A}_{S}={\mbox {STr}}\ln (\hat{Q}_{0}^{-1}-\beta ^{2}\hat{Q})
\label{A_S_simple}
\end{equation}
where $\mbox {STr}$ denotes integral over times and trace over
supersymmetric variables, $\widehat{Q}$ are understood as operators and $%
\hat{Q}_{0}=\langle \phi ({\cal T}_{1})\phi ({\cal T}_{2})\rangle $.
Explicitly $\hat{Q}_{0}^{(B)}=D_{t_{1},t_{2}}^{0}+G_{t_{1},t_{2}}^{0}\bar{%
\theta _{2}}\theta _{2}+G_{t_{2},t_{1}}^{0}\bar{\theta _{1}}\theta _{1}$.
Here $G_{0}$ and $D_{0}$ are single site spin response and correlation
functions which contain all local effects. Inverting $Q_{0}$ we get $(\hat{Q}%
_{0}^{(B)})^{-1}=-(G_{0}^{-2}D_{0})_{t_{1},t_{2}}^{{}}+(G_{0}^{-1})_{t_{1},t_{2}}^{{}}%
\bar{\theta _{2}}\theta _{2}+(G_{0}^{-1})_{t_{2},t_{1}}^{{}}\bar{\theta _{1}}%
\theta _{1}$, next we insert it into (\ref{A_S_simple}) and rewrite it in a
more explicit form: 
\[
{\cal A}_{S}=\frac{1}{2}\ln \det \left| 
\begin{array}{ll}
-\beta ^{2}\widehat{D} & \left( G_{0}^{-1}-\beta ^{2}G\right)  \\ 
\left( G_{0}^{-1}-\beta ^{2}G\right)  & -(G_{0}^{-2}D_{0})-\beta ^{2}D
\end{array}
\right| 
\]

Finally, as we shall see below, transitions over the barriers in the
vicinity of $T_{c}$ require only small noise fields $\widehat{D}$; this
allows us to simplify the action further by expanding in $\widehat{D}$;
retaining only the leading and subleading terms in it we get the simplified
expression for the full action (\ref{A}):

\begin{eqnarray}
{\cal A} &=&\frac{1}{2}\beta ^{2}{\mbox {Tr}}\left[ GG+D\widehat{D}\right] +{%
\mbox {Tr}}\ln \left( G_{0}^{-1}-\beta ^{2}G\right) -  \label{A_1} \\
&&\frac{1}{2}\beta ^{2}{\mbox {Tr}}\left( G_{0}^{-1}-\beta ^{2}G\right)
^{-1}\Pi \left[ \left( G_{0}^{-1}-\beta ^{2}G^{-1}\right) ^{\dagger }\right]
^{-1}\widehat{D}  \nonumber
\end{eqnarray}
where $G$, $D$, $\widehat{D}$ and $\Pi \equiv (G_{0}^{-2}D_{0})+\beta ^{2}D$
are understood as operators so that, e.g. ${\mbox {Tr}}GG{\equiv }\int
dt_{1}dt_{2}G_{t_{1}t_{2}}G_{t_{2}t_{1}}$. Below we shall get an equation
for $D$ by varying this action with respect to $\widehat{D}$, so to get the
leading and subleading terms in this equation we shall need to keep also
terms up to quadratic order in $\widehat{D}$: 
\begin{eqnarray}
{\cal A} &=&\frac{1}{2}\beta ^{2}{\mbox {Tr}}\left[ GG+D\widehat{D}\right] +{%
\mbox {Tr}}\ln \left( G_{0}^{-1}-\beta ^{2}G\right) -  \label{A_simple_quad}
\\
&&\frac{1}{2}\beta ^{2}{\mbox {Tr}}\left[ G_{0}^{-1}-\beta ^{2}G\right]
^{-1}\Pi \left[ (G_{0}^{-1}-\beta ^{2}G)^{\dagger }\right] ^{-1}\widehat{D}-
\nonumber \\
&&\frac{1}{4}\beta ^{4}{\mbox {Tr}}\left\{ \left[ G_{0}^{-1}-\beta
^{2}G\right] ^{-1}\Pi \left[ (G_{0}^{-1}-\beta ^{2}G)^{\dagger }\right] ^{-1}%
\widehat{D}\right\} ^{2}  \nonumber
\end{eqnarray}
Varying action (\ref{A_simple_quad}) with respect to $G$, $\widehat{D}$ and $%
D$ fields and keeping only the leading and subleading terms in $\widehat{D}$
we get equations 
\begin{eqnarray}
G &=&\left( G_{0}^{-1}-\beta ^{2}G\right) ^{-1}+G\Pi G^{\dagger }\hat{D}G
\label{G_simple} \\
D &=&G\Pi G^{\dagger }-G\Pi G^{\dagger }\widehat{D}G\Pi G^{\dagger }
\label{D_simple} \\
\widehat{D} &=&G^{\dagger }\widehat{D}G-G^{\dagger }\widehat{D}G\Pi
G^{\dagger }\widehat{D}G  \label{Dhat_simple}
\end{eqnarray}

Here we have dropped the factors of $\beta $ in the subleading terms in $%
\widehat{D}$, ignoring their dependence on $\tau =T_{c}-T$ because these
terms are already small in $\tau $ as explained above; we simplified the
second terms in all the equations noting that they are already of the next
order in $\widehat{D}$,  which allows us to substitute $\left[
G_{0}^{-1}-\beta ^{2}G\right] ^{-1}\approx G$ in these terms; finally,
we have also simplified the second and the third equations using the first
equation to express $\left( G_{0}^{-1}-\beta ^{2}G\right) ^{-1}$ via $G$.
These equations are analogues of the equations (\ref{x_eq}) for the toy
model. At $\widehat{D}=0$ this system of equations reduces to the equations
derived in \cite{Sompolinsky82} for the glass dynamics above $T_{c}$,
leading to the response and correlation functions that satisfy
fluctuation-dissipation theorem (FDT). As was shown in \cite{Sompolinsky82}
the quadratic approximation to the local spin response is sufficient above $%
T_{c}$ so in this temperature range the equations (\ref{G_simple}-\ref
{Dhat_simple}) are correct. However, in this temperature range there are no
barriers and so for our present purpose these equations are not very
interesting either. To finish our discussion of the equations for $T>T_{c}$
we give the final formula for the action expressed via \thinspace the
solutions $G$, $\widehat{D}$ and $D$ of the equations (\ref{G_simple}-\ref
{Dhat_simple}) which are simplified by making the same approximations as we
did in deriving equations (\ref{G_simple}-\ref{Dhat_simple}): 
\begin{eqnarray}
{\cal A} &=&\frac{1}{2}\beta ^{2}{\mbox {Tr}}\left[ GG+D\widehat{D}\right] +{%
\mbox {Tr}}\ln \left( G_{0}^{-1}-\beta ^{2}G\right) -  \label{A_simple_final}
\\
&&\frac{1}{2}\beta ^{2}{\mbox {Tr}}G\Pi G^{\dagger }\widehat{D}+\frac{3}{4}%
\beta ^{4}{\mbox {Tr}}\left\{ G\Pi G^{\dagger }\widehat{D}\right\} ^{2} 
\nonumber
\end{eqnarray}
Note that because we have already used the equations (\ref{G_simple}-\ref
{Dhat_simple}) to simplify the action the solutions of these equation do not
necessarily minimize the action (\ref{A_simple_final}).

We now derive the full equations below $T_{c}$ where non-linearity of the
action ${\cal A}_{0}\{\phi \}$ becomes essential \cite{Sompolinsky82},
employing a diagram technique. To construct this technique we note that the
saddle point equations can be rewritten in the form $\hat{D}%
_{t_{1},t_{2}}=\langle \hat{S}_{t_{1}}\hat{S}_{t_{2}}\rangle _{S}$, $%
D_{t_{1},t_{2}}=\langle S_{t_{1}}S_{t_{2}}\rangle _{S}$ and $%
G_{t_{1},t_{2}}=\langle S_{t_{1}}\hat{S}_{t_{2}}\rangle _{S}$ where $\langle
\dots \rangle _{S}$ denotes the average over a single spin variable taken
with the weight $\exp (-{\cal A}_{0}\{\phi \}+\int Q_{{\cal T}_{1},{\cal T}%
_{2}}\phi ({\cal T}_{1})\phi ({\cal T}_{2})d{\cal T}_{1}d{\cal T}_{2})$. We
shall use a formal expansion in the term $\int Q_{{\cal T}_{1},{\cal T}%
_{2}}\phi ({\cal T}_{1})\phi ({\cal T}_{2})d{\cal T}_{1}d{\cal T}_{2}$ to
derive equations for the correlation and response functions and then
reconstruct the action that gives them; furthermore, as above, we shall keep
only the leading and subleading terms in $\widehat{D}$ in these equations.

The known form of the equations for $D_{t_1,t_2}$ and $G_{t_1>t_2}=0$ in the
case when $\hat{D}_{t_1,t_2}=0$ and $G_{t_1<t_2}=0$ allows us to reconstruct
all the terms in the action ${\cal A}$ that are linear in the conjugate
fields. We shall go briefly over the derivation of these equations because
later it will allow us more easily to explain how to augment these equations
for non-zero conjugate fields and construct the full action.

We start with the equation for the response function $G$ at $\widehat{D}=0$.
In this approximation $G$ satisfies the Dyson equation diagrammtically
represented as shown in Figs 1a and Fig 1b, the first of these contributions
sufficing to reproduce the correct dynamics above $T_{c}$ and  equivalent to
(\ref{G_simple}) at $\widehat{D}=0$ while the second is needed to reproduce
replica symmetry breaking effects in thermodynamics and memory effects in
the correct dynamical solution \cite{SGbooks,Sompolinsky82,Kurchan93}.
Together they yield 
\begin{equation}
G=[G_{0}^{-1}-\beta G\beta -3y(GD^{2})]^{-1},  \label{G_ad}
\end{equation}
which leads to the conventional equation for the dynamics below $T_{c}$ and
eventually to the result (\ref{q_ad}). Here and below inversion should be
understood as operator inversion, $\beta G\beta $ stands for $\beta
_{t_{1}}G_{t_{1},t_{2}}\beta _{t_{2}}$, $G_{0}$ is the bare local spin
response function and we use parenthesis $\left( {}\right) $ to imply the
usual (arithmetic) product of two functions as opposed to the operator
product implied everywhere else. The exact form of $G_{0}^{-1}$ is not
important, because we need only its low frequency asymptote for which we
assume a general form $G_{0}(\omega )^{-1}=a-i\omega \Gamma $ where $a$ is
the renormalized zero frequency response and $\Gamma $ is a renormalized
relaxation rate; these renormalizations are due to high energy processes and
are not singular. Finally, the coefficient $y$ describes the strength of a
local four spin correlator, in the following we shall take the value $y=2/3$
corresponding to the hard Ising spins \cite{Sompolinsky82,Ioffe88,Kurchan93}%
, but note that this value can be always reduced to unity by scaling the
correlation functions in the final slow cooling equations, so it is not
important for the qualitative properties of the solutions. The factor $3$ in
front of the subleading term in Eq. (\ref{G_ad}) is a combinatorial factor
associated with the self energy diagram, it appears because two intermediate
propagators in it are $D$s and only one is $G$; note that it does not appear
in the corresponding diagram for $D$ where all intermediate propagators are $%
D$s. A non-zero conjugate field $\widehat{D}$ gives a new contribution to
the equation for $G$. Diagrammatically this is as shown in Fig. 1a, 1b
(terms giving Eq. (\ref{G_ad})) and Fig. 1c (new term). Combining and
manipulating these terms leads to 
\begin{eqnarray}
G &=&\widetilde{G}+G\Pi G^{\dagger }\hat{D}G  \label{G_eq} \\
\widetilde{G} &\equiv &\left[ G_{0}^{-1}-\beta G\beta -3y(D^{2}G)\right]
^{-1}  \label{Gtilde} \\
\Pi  &\equiv &G_{0}^{-2}D_{0}+\beta D\beta +y(D^{3})  \label{Pi}
\end{eqnarray}
where we have dropped factors of $\beta \approx 1$ in the terms subleading
in $\widehat{D}$ because the corrections are small in the vicinity of $T_{c}$
and have replaced $\widetilde{G}$ by $G$ in the second term because the
difference between these is $O(\widehat{D})$.

The equation for the correlation function $D_{t_{1},t_{2}}$ in the leading
approximation in $\widehat{D}$ is shown in Fig. 1d; in analogy with the
equation for the response function shown in Fig 1a we keep here only the
leading and subleading terms in the self energy. This equation together with
the equation for $G$ yields the full system of equations for the dynamics of
the spin glass below the transition temperature in the absence of 
transitions over the barriers; the adiabatic approximation to these
equations was discussed in e.g. \cite{Vinokur87,Ioffe88}. The graphical
representation of the subleading term in $\widehat{D}$ is shown in Fig. 1e;
it is analogous to the subleading term in the equation (\ref{D_simple})
obtained in the quadratic approximation. Combining these terms we get 
\begin{equation}
D=\widetilde{G}\Pi \widetilde{G}^{\dagger }+\widetilde{G}\Pi \widetilde{G}%
^{\dagger }\hat{D}\widetilde{G}\Pi \widetilde{G}^{\dagger }  \label{D_eq_1}
\end{equation}
Again, to the order required we may replace $\widetilde{G}$ by $G$ in the
second term. \ A further simplifiaction results from using (\ref{G_eq}) to
express $\widetilde{G}$ via $G$ in the first term yielding

\begin{equation}
D=G\Pi G^{\dagger }-G\Pi G^{\dagger }\hat{D}G\Pi G^{\dagger }  \label{D_eq}
\end{equation}
Finally, the equation for $\widehat{D}$ follows from the summation of the
diagrams shown in Fig. 1f and Fig. 1g; note that here we have two
contributions to the subleading part that graphically differ by the
directions of the arrows inside the self energy term:

\begin{eqnarray}
\widehat{D} &=&G^{\dagger }\left[ \beta \widehat{D}\beta +3y(D^{2}\widehat{D}%
)+6y(GG^{\dagger }D)\right] G  \label{Dhat_eq} \\
&&-G^{\dagger }\hat{D}G\Pi G^{\dagger }\hat{D}G  \nonumber
\end{eqnarray}
Here, as above, we have simplified the final expression using equations (\ref
{G_eq}) to express $\widetilde{G}$ via $G$ in the first term, replaced $%
\widetilde{G}$ by $G$ in the second term and neglected factors of $\beta
\approx 1$ and terms proportional to $y\left[ (D^{2}\widehat{D}%
)+(GG^{\dagger }D)\right] \widehat{D}$ because all these effects are small
in the vicinity of $T_{c}$.

Equations (\ref{G_eq}-\ref{Dhat_eq}) form the full system of equations
describing dynamics of the spin glass below $T_{c}$, it differs from the
system (\ref{G_simple}-\ref{Dhat_simple}) (that can be used above $T_{c}$)
only by the terms proportional to $y$ which describe the strength of the
local non-linear spin response. The action corresponding to the system of
equations (\ref{G_eq}-\ref{Dhat_eq}) is rather cumbersome, so we shall not
write it here. Instead we shall first further simplify these equations by
assuming that all fields change adiabatically slowly, i.e. that the time
scale at which they change is much longer than the spin flip time scale $%
\Gamma ^{-1}$. In this approximation the fast part of the Green function can
be replaced by a delta-function, so we get $G_{t_{1},t_{2}}=(1-q)\delta
(t_{1}-t_{2})+\Delta _{t_{1},t_{2}}$. We shall simplify these equations even
more by keeping only the leading terms in the reduced temperature $\tau $
and using the fact that $\tau -q\equiv \delta q=O(\tau ^{2})$. To prove this
general statement consider, say, equation (\ref{D_eq}) at $t_{1}-t_{2}\gg
\Gamma _{0}^{-1}$ and keep only the terms of the order of $\tau ^{2}$. The
second term in this equation is at least of the order of $\tau ^{3}$ and can
be ignored completely. Moreover in the approximation one can replace $%
G_{t_{1},t_{2}}\rightarrow (1-q)\delta (t_{1}-t_{2})$ and $\Pi \rightarrow
\beta D\beta $ in the first term. Collecting the remaining terms we get $%
(\tau -q)D_{t_{1},t_{2}}=O(\tau ^{3})$, so the existence of a non-zero $%
D_{t_{1},t_{2}}=q_{t_{1},t_{2}}=O(q)$ implies that $\delta q=O(\tau ^{2})$.
The remaining two equations are also satisfied to the order of $\tau ^{2}$
if $(\tau -q)D_{t_{1},t_{2}}=O(\tau ^{3})$, so one gets non-trivial
equations keeping only the terms of the order of $\tau ^{3}$: 
\begin{eqnarray}
&&
\begin{array}{ll}
(\delta q_{t_{1}}+\delta q_{t_{2}}+3yq_{t_{1},t_{2}}^{2})\Delta
_{t_{1},t_{2}}+ &  \\ 
\int (\Delta _{t_{1},t}\Delta _{t,t_{2}}\!+\!q_{t_{1},t}\hat{D}%
_{t,t_{2}}\!)dt=0 & 
\end{array}
\label{Delta_eq} \\
&&
\begin{array}{ll}
(\delta q_{t_{1}}+\delta q_{t_{2}}+yq_{t_{1},t_{2}}^{2})q_{t_{1},t_{2}}+\int
q_{t_{1},t}\Delta _{t_{2},t}\!dt+ &  \\ 
\int q_{t_{2},t}\Delta _{t_{1},t}-\int q_{t_{1},t^{\prime }}\hat{D}%
_{t^{\prime },t^{\prime \prime }}q_{t^{\prime \prime },t}dt^{\prime
}dt^{\prime \prime }=0 & 
\end{array}
\label{q_eq} \\
&&
\begin{array}{l}
(\delta q_{t_{1}}+\delta q_{t_{2}}+3yq_{t_{1},t_{2}}^{2})\hat{D}%
_{t_{1},t_{2}}+ \\ 
\int \hat{D}_{t_{1},t}\Delta _{t,t_{2}}\!dt+\!\!\int \hat{D}_{t,t_{2}}\Delta
_{t,t_{1}}dt+ \\ 
6yq_{t_{1},t_{2}}\Delta _{t_{1},t_{2}}\Delta _{t_{2},t_{1}}-\int \widehat{D}%
_{t_{1},t^{\prime }}q_{t^{\prime },t^{\prime \prime }}\widehat{D}_{t^{\prime
\prime },t}dt^{\prime }dt^{\prime \prime }=0
\end{array}
\label{qhat_eq}
\end{eqnarray}

Now we derive the simplified expression for the action that corresponds to
the adiabatic and $\tau \ll 1$ approximations used in deriving (\ref
{Delta_eq}-\ref{qhat_eq}). First we note that the spin part of the action, $%
{\cal A}_{S}$, is zero if $\hat{D}\equiv 0$, $\Delta _{t_{1}<t_{2}}\equiv 0$%
; this can most easily be verified by inspecting the diagrammatic expansion
for the action. In this representation ${\cal A}_{S}$ is given by the sum of
closed diagrams which contain only $G$ lines or contain at least one $\hat{D}
$ line so at least one of the of the lines in any closed diagram is either $%
\Delta _{t_{1}<t_{2}}$ or $\hat{D}$, therefore ${\cal A}_{S}=0$ if both
these functions are zero. Further, variational derivatives of the action
with respect to the functions $\hat{D}$ and $G$ give spin correlators, i.e. $%
-\frac{\delta {\cal A}_{S}}{\beta ^{2}\delta G}=\widetilde{G}$ $\equiv
\left[ G_{0}^{-1}-\beta G\beta -3y(D^{2}G)\right] ^{-1}$ at $\hat{D}\equiv 0$
and $-\frac{\delta {\cal A}_{S}}{\delta \widehat{D}}=\frac{1}{2}\left[ 
\widetilde{G}\Pi \widetilde{G}^{\dagger }+\widetilde{G}\Pi \widetilde{G}%
^{\dagger }\hat{D}\widetilde{G}\Pi \widetilde{G}^{\dagger }\right] $ so
these two equations completely determine it. To simplify the action in the
vicinity of the transition temperature we note that the part of the action
which is zeroth order in $\hat{D}$ necessarily contains at least one
advanced response function, $G_{A}$. As discussed above, we expect that in
the vicinity of $T_{c}$ such anomalous terms are small (because barriers are
small) and so we can expand in $\hat{D}$. Keeping at most terms of the
second order in $G_{A}$ and $\hat{D}$ we get

\begin{eqnarray}
{\cal A}_{S} &=&-\beta ^{2}{TrG}_{A}\widetilde{G}_{R}-\frac{1}{2}{\mbox {Tr}}%
\left[ {G}_{A}\widetilde{G}_{R}\right] ^{2}-  \label{A_S_expanded} \\
&&\frac{1}{2}{\mbox {Tr}}\left[ \widetilde{G}\Pi \widetilde{G}^{\dagger }%
\hat{D}+\frac{1}{2}\left[ \widetilde{G}\Pi \widetilde{G}^{\dagger }\hat{D}%
\right] ^{2}\right]   \nonumber \\
\widetilde{G}_{R} &\equiv &\left[ G_{0}^{-1}-\beta G_{R}\beta
-3y(D^{2}G_{R})\right] ^{-1}  \nonumber
\end{eqnarray}
where in analogy with the derivation of the equations (\ref{G_eq},\ref{D_eq},%
\ref{Dhat_eq}) we have neglected self energy corrections to the terms that
are of the second order in $G_{A}$. The full action is ${\cal A}={\cal A}%
_{S}+\beta ^{2}\left[ {\mbox {Tr}}G_{A}G_{R}+\frac{1}{2}{\mbox {Tr}}D%
\widehat{D}\right] $. We can simplify this expression by replacing $G_{R}$
and $D$ in the term $\beta ^{2}\left[ {\mbox {Tr}}G_{A}G_{R}+\frac{1}{2}{%
\mbox {Tr}}D\widehat{D}\right] $ by the r.h.s of the equations (\ref{G_eq},%
\ref{D_eq}). Keeping only the terms of the second order in $G_{A}$ and $\hat{%
D}$ and retaining $\beta $ only in the leading terms we get 
\begin{eqnarray*}
{\cal A} &=&\frac{1}{2}{\mbox {Tr}}\left[ {G}_{A}G_{R}\right] ^{2}+{%
\mbox
{Tr}}\left[ G_{A}G_{R}\Pi G_{R}\widehat{D}G_{R}\right] - \\
&&\frac{1}{4}{\mbox {Tr}}\left[ {G\Pi G}^{\dagger }\widehat{D}\right] ^{2}
\end{eqnarray*}
Finally we assume that $G_{A}$ and $\hat{D}$ are adiabatically slow and keep
only the leading terms in $\tau $:

\begin{equation}
{\cal A}={\mbox {Tr}}\left[ \Delta _{A}\Delta _{A}\Delta _{R}+\Delta _{A}%
\hat{D}q-\frac{1}{4}q\hat{D}q\hat{D}\right]  \label{A_final}
\end{equation}
where $\Delta _{A}\equiv \Delta _{t_{1}t_{2}}\theta (t_{2}-t_{1})$ ($\Delta
_{R}\equiv \Delta _{t_{1}t_{2}}\theta (t_{1}-t_{2})$) are advanced
(retarded) parts of the adiabatic response.

For the numerical solution of the equations (\ref{Delta_eq}-\ref{qhat_eq})
it is convenient to regard them as the equations for the saddle point of the
effective action 
\begin{eqnarray}
{\cal A}_{eff} &=&-\frac{1}{3}\int \Delta _{t_{1},t_{2}}\Delta
_{t_{2},t_{3}}\Delta _{t_{3},t_{1}}dt_{1}dt_{2}dt_{3}-  \label{A_eff} \\
&&\frac{1}{2}\int (\delta q_{t_{1}}+\delta
q_{t_{2}}+yq_{t_{1},t_{2}}^{2})q_{t_{1},t_{2}}\widehat{D}%
_{t_{2},t_{1}}dt_{1}dt_{2}-  \nonumber \\
&&\frac{1}{2}\int (\delta q_{t_{1}}+\delta
q_{t_{2}}+3yq_{t_{1},t_{2}}^{2})\Delta _{t_{1},t_{2}}\Delta
_{t_{2},t_{1}}dt_{1}dt_{2}-  \nonumber \\
&&\int q_{t_{1},t_{2}}\widehat{D}_{t_{2},t_{3}}\Delta
_{t_{3},t_{1}}dt_{1}dt_{2}dt_{3}+  \nonumber \\
&&\frac{1}{4}\int q_{t_{1},t_{2}}\widehat{D}_{t_{2},t_{3}}q_{t_{3},t_{4}}%
\widehat{D}_{t_{4},t_{1}}dt_{1}dt_{2}dt_{3}dt_{4}  \nonumber
\end{eqnarray}
Using the equations (\ref{Delta_eq}-\ref{qhat_eq}) and representing $\Delta $
as a sum of the retarded and advanced parts we can simplify the action (\ref
{A_eff}) and get the same result (\ref{A_final}) as above.

Scaling analysis of the equations (\ref{Delta_eq}-\ref{qhat_eq}) shows that
their solutions have the following scales: $\Delta \sim \tau \frac{d\tau }{dt%
}$, $q\sim \tau $ and $\hat{D}\sim \tau (\frac{d\tau }{dt})^{2}$. Using
these values to estimate the last (quadratic in $\hat{D}$) term in (\ref
{A_final}) we see that it scales as $\int (\tau (\frac{d\tau }{dt}%
)^{2})^{2}\tau ^{2}dt^{4}\propto \tau ^{8}$ whereas the first two terms (and
the action itself scale as $\int (\tau (\frac{d\tau }{dt}))^{3}dt^{3}\propto
\tau ^{6}$ and $\int \tau (\frac{d\tau }{dt})\tau (\frac{d\tau }{dt}%
)^{2}\tau dt^{3}\propto \tau ^{6}$. These estimates allow us to neglect the
last terms in the expressions (\ref{A_final}) and (\ref{A_eff}) for the
action and the last terms in the equations (\ref{q_eq}) and (\ref{qhat_eq}).

We were not able to guess the ansatz for the analytical solution of the
equations (\ref{Delta_eq}-\ref{qhat_eq}). Instead we show numerically that
the solution that we expected on physical grounds indeed exists. As
explained above, we expect that as the temperature is decreased a given
metastable state continues to subdivide and that the barriers between these
filial states are smaller than the barriers separating the ancestor states.
Therefore we expected that a rare fluctuation (``instanton'') might take the
system from one of these filial states to another leaving the ancestor state
unchanged. In this case the memory of the perturbation applied at the times
when these filial states were formed is completely lost, i.e. $\Delta
_{t_{1},t_{2}}=0$ for $t_{2}$ greater than some $t_{*}$. Indeed, we observe
a solution with this property; in Fig 2a we show $\Delta _{t_{1},t_{2}}$
obtained numerically for the slow cooling process which leads to the final
reduced temperature $\tau _{f}=0.1$. This numerical solution clearly has the
property that the memory of the latest perturbation is erased. Note also
that the conjugate field corresponding to this solution is small: it is
proportional to $\tau $ and moreover seems to have a numerically small
coefficient. This solution corresponds to the action $S/N\approx 0.1\tau ^{6}
$. By construction this solution is not time-invariant, instead it is
localized at the end of the cooling procedure. We expect that these
equations also admit a solution of a different kind which occurs when the
system is first slowly cooled down to the final temperature, $\tau
(t_{f})=\tau _{f}$, and then kept at this temperature for a long time. This
solution corresponds to a transition over the barrier at some $t_{t}$,  in
this solution $\Delta _{t_{1},t_{2}}=0$ for $t_{*}<t_{2}<t_{f}$ and $%
t_{t}<t_{1}$. Such a solution should be invariant under a shift of $t_{2}$,
but, unfortunately, we could not study such solutions numerically because of
a much larger number of time points required for such study. We expect that
solutions with $t_{*}\rightarrow t_{f}$ correspond to the transition between
the states on the lowest level of the hierarchy and erase only a memory of
the events in the narrow time range $(t_{*},t_{f})$ when this hierarchy was
formed; from (\ref{A_final}) it is clear that such solutions have very small
action (suppressed by additional power of $(\tau _{*}-\tau _{f})^{2}$), so
there is no lower bound on the energy of the barrier in this approach. This
low bound might appear when terms of next order in $1/N$ are taken into
account and it is quite probable that the numerical result \cite{Mackenzie82}
that the lowest barriers scale with $1/N^{1/4}$ is due to this mechanism.

In conclusion we have developed the formalism which allows one to study the
distribution of barriers in a spin glass with a large but finite number of
spins, $N$. We considered the particular example of the
Sherrington-Kirkpatrick model and derived analytically the equations for the
dynamic order parameters of this model that admit transition over the
barriers. We also showed numerically that these equations admit a
non-trivial solution for which the energy of the barrier scales as $N\tau
^{6}$. It remains an open question whether this solution is unique or in
fact these equations admit a whole family of solutions and the energy of the
barriers has a broad distribution. It also remains to be investigated
whether these results and general approach can be applied to the common
physical situation of a spin glass where each spin has finite number of
neighbours but the total number of spins is infinite.

{\bf Acknowledgement:} One of us (LI) would like to thank the Physics
Departement and New College at Oxford University for hospitality during the
execution of this research. Financial support of the UK EPSRC under grant DR
8729 is also gratefully acknowledged.

\newpage

\centerline{\epsfxsize=8.5cm \epsfbox{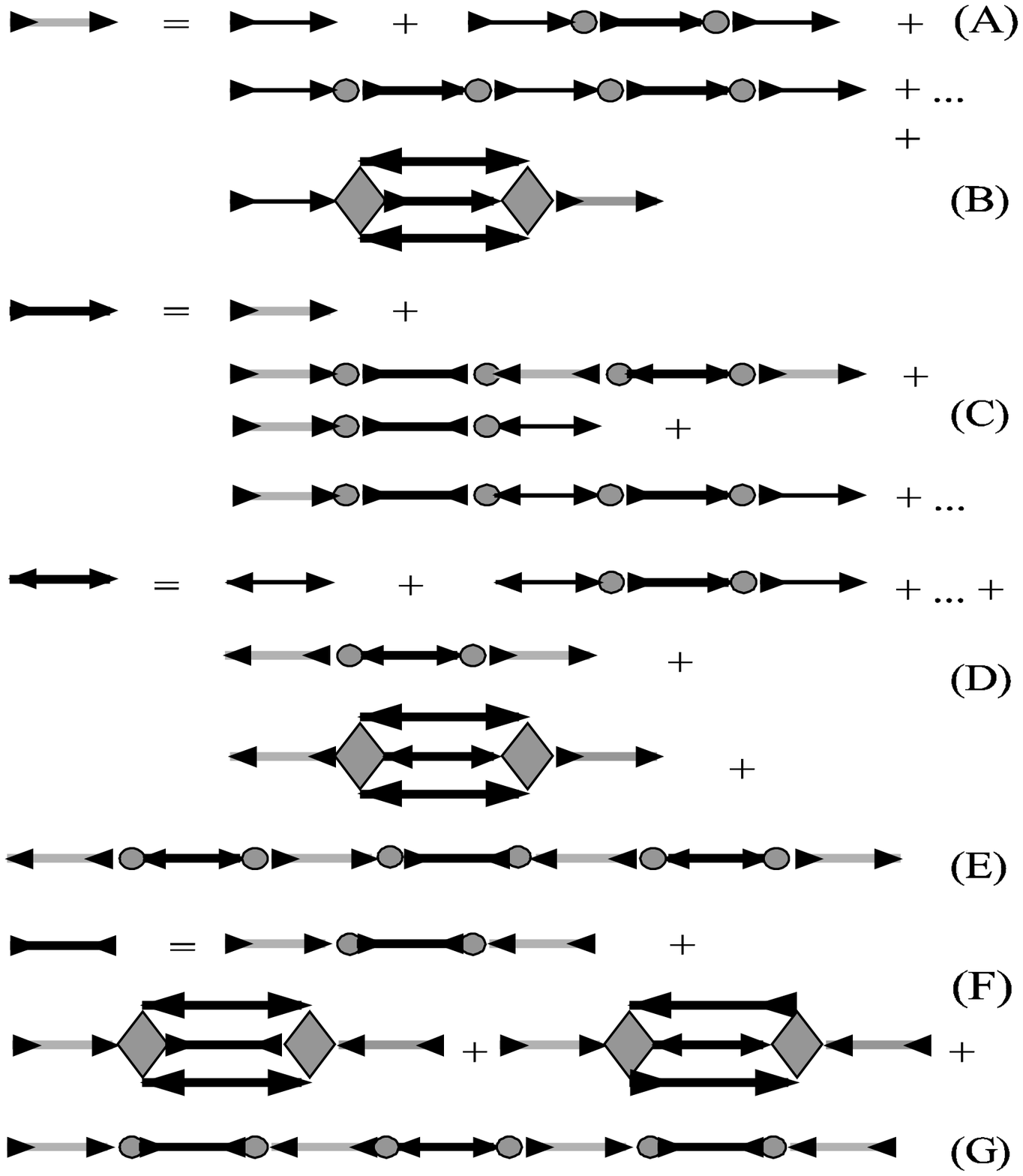}}

{\footnotesize {\bf Fig. 1} Diagrams giving equations (\ref{G_eq}-\ref
{Dhat_eq}). Here the lines with the arrows in the same direction denote $G$,
the lines with arrows pointing outwards denote $D$ and the lines with arrows
pointing inwards denote $\hat{D}$. Thick lines are full (renormalized)
correlation functions, a thin line with arrows in the same direction is a
bare Green function $G_{0}$, a gray thick line is $\widetilde{G}$ i.e. the
response function calculated in the zeroth order in $\hat{D}$, a gray circle
denotes $\beta $ and a gray diamond denotes a four spin vertex, $\sqrt{y}$%
.The geometric series for $\widetilde{G}$ is shown in (a) and (b); this
series sums to (\ref{Gtilde}) which is equivalent to the well known \cite
{Sompolinsky82,Vinokur87,Ioffe88} dynamical equations. First order
corrections to $G$ are shown in (c); as is clear from this series a general
diagram contains $0,1,\dots $ bare Green functions to the right of $D_{0}$
and $0,1,\dots $ between $\hat{D}$ and $D_{0}$ so the sum of these terms is $%
\widetilde{G}G_{0}^{-1}$ ($G_{0}^{-1}\widetilde{G}$) making the sum of all
diagrams linear in $\hat{D}\widetilde{G}\Pi \widetilde{G}^{\dagger }\widehat{%
D}\widetilde{G}$ with $\Pi $ that is given by (\ref{Pi}). In (d) we show the
leading terms in the expansion for $D$, this series is similar to the one
shown in (c) and gives $\widetilde{G}\Pi \widetilde{G}^{\dagger }$. In (e)
we show one term which gives the first order correction to $D$; this term is
proportional to $D^{2}$ and together with the analogous terms that are
proportional to $DD_{0}$ or $D_{0}^{2}$ it gives $\widetilde{G}\Pi 
\widetilde{G}^{\dagger }\widehat{D}\widetilde{G}\Pi \widetilde{G}^{\dagger }.
$ Finally, in (f) and (g) we show the leading and subleading terms for $\hat{%
D}$; note that the leading term for this function does not contain a bare
part and moreover its self-energy contains either $\hat{D}$ or two response
functions with opposite directions ensuring that the usual solution (in
which $\hat{D}=0$ and response is purely retarded) is self-consistent.}

\centerline{\epsfxsize=8.5cm \epsfbox{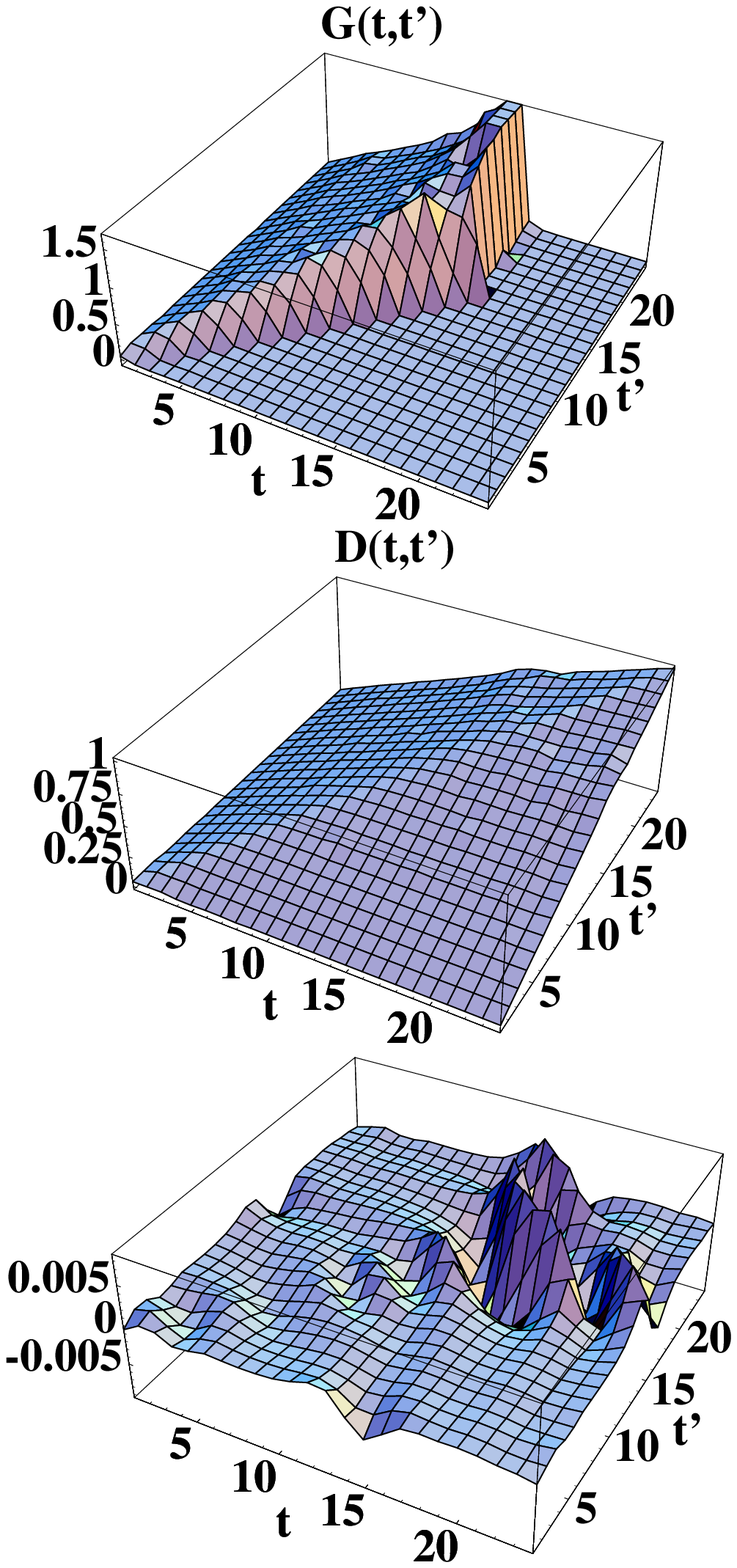}}

{\footnotesize {\bf Fig. 2} A numerical solution of the system of equations (%
\ref{Delta_eq}-\ref{qhat_eq}) for the slow cooling process leading to the
final reduced temperature $\tau (t_{f})=\tau _{f}=0.1$. All displayed
functions are scaled by $\tau _{f}$. Fig 2a shows $\Delta $, Fig 2b shows $D$
and Fig. 2c shows $\hat{D}$. From Fig 2a it is clear that
in this solution a perturbation applied during the later part of the cooling
process, $t_{tr}<t<t_{f}$  has no effect on the state at $%
t_{f}$, so this solution corresponds to the transition that erases memory of
these times; we also observe that this transition has relatively little
effect on the correlation function $D$ but results in the
appearance of the 'noise' field $\hat{D}$ localized around times $t\sim
t_{tr}$.}

\end{multicols}

\end{document}